\documentclass[reprint,amsmath,amssymb,aps,prc,showpacs,longbibliography,lengthcheck,]{revtex4-1}

\usepackage{graphicx}
\usepackage{epsfig}
\usepackage{color}
\usepackage{multirow}

\usepackage{amsmath}
\usepackage{amssymb}
\usepackage{bm}
\long\def\symbolfootnote[#1]#2{\begingroup%
\def\thefootnote{\fnsymbol{footnote}}\footnote[#1]{#2}\endgroup}
\begin{document}


\title{Measurement of the $\Sigma$ beam asymmetry for the $\omega$ photo-production off the proton and the neutron at GRAAL}
\author{V.~Vegna$^{1,2}$}
	\altaffiliation{Present affiliation: Physikalisches Institut - Bonn Universit\"{a}t, Nussallee 12, D-53115 Bonn, Deutschland;\\email: vegna@physik.uni-bonn.de}
\author{A.~D'Angelo$^{1,2}$}
\author{O.~Bartalini$^{1,2}$}
\author{V.~Bellini$^{3,4}$}
\author{J.-P.~Bocquet$^5$}
\author{M.~Capogni$^{1,2}$} \altaffiliation{Present affiliation: ENEA - C.R. Casaccia, via Anguillarese 301, I-00060 Roma, Italy}
\author{L.E.~Casano$^2$}
\author{M.~Castoldi$^6$}
\author{F.~Curciarello$^{7,4}$} 
\author{V.~De Leo$^{7,4}$}
\author{J.-P.~Didelez$^8$}
\author{R.~Di Salvo$^2$}
\author{A.~Fantini$^{1,2}$}
\author{D.~Franco$^{1,2}$} \altaffiliation{Present affiliation: IPNL - 43, Bd du 11 Novembre 1918, Fr69622 Villeurbanne Cedex, France}
\author{G.~Gervino$^{9,10}$}
\author{F.~Ghio$^{11,12}$}
\author{G.~Giardina$^{7,4}$}
\author{B.~Girolami$^{11,12}$}
\author{A.~Giusa$^{3,4}$}
\author{A.~Lapik$^{13}$}
\author{P.~Levi Sandri$^{14}$}
\author{A.~Lleres$^5$}
\author{F.~Mammoliti$^{3,4}$}
\author{G.~Mandaglio$^{7,4,15}$}
\author{M.~Manganaro$^{7,4}$} \altaffiliation{Present affiliation: Tyndall National Institute - University College Cork, Dyke Parade, Cork, Ireland}
\author{D.~Moricciani$^2$}
\author{A.~Mushkarenkov$^{13}$}
\author{V.~Nedorezov$^{13}$}
\author{C.~Randieri$^{3,4}$}
\author{D.~Rebreyend$^5$}
\author{N.~Rudnev$^{13}$}
\author{G.~Russo$^{3,4}$}
\author{C.~Schaerf$^{1,2}$} 
\author{M.-L.~Sperduto$^{3,4}$}
\author{M.-C.~Sutera$^4$}
\author{A.~Turinge$^{13}$}
\author{I.~Zonta$^{1,2}$}

\affiliation{$^1$Dipartimento di Fisica - Universit\`{a} degli Studi di Roma ``Tor Vergata'', via della Ricerca Scientifica 1, I-00133 Roma, Italy}
\affiliation{$^2$INFN - Sezione di Roma ``Tor Vergata'', via della Ricerca Scientifica 1, I-00133 Roma, Italy}
\affiliation{$^3$Dipartimento di Fisica - Universit\`{a} degli Studi di Catania, via Santa Sofia 64, I-95123 Catania, Italy}
\affiliation{$^4$INFN - Sezione di Catania, via Santa Sofia 64, I-95123 Catania, Italy}
\affiliation{$^5$LPSC, Universit \'{e} Joseph Fourier Grenoble1, CNRS/IN2P3, Grenoble IPN, F-38026 Grenoble, France}
\affiliation{$^6$Dipartimento di Fisica - Universit\`{a} degli Studi di Genova, via Dodecaneso 33, I-16146 Genova, Italy}
\affiliation{$^{7}$Dipartimento di Fisica e di Scienze della Terra, Universit\`{a} di Messina, salita Sperone 31, I-98166 Messina, Italy}
\affiliation{$^8$IN2P3, Institut de Physique Nucl\'{e}aire, Rue Georges Clemenceau, F-91406 Orsay, France}
\affiliation{$^9$Dipartimento di Fisica Sperimentale - Universit\`{a} degli Studi di Torino, via Pietro Giuria 1, I-10125 Torino, Italy}
\affiliation{$^{10}$INFN - Sezione di Torino, via Pietro Giuria 1, I-10125 Torino, Italy}
\affiliation{$^{11}$Istituto Superiore di Sanit\`{a}, viale Regina Elena 299, I-00161 Roma, Italy}
\affiliation{$^{12}$INFN - Sezione di Roma, piazzale Aldo Moro 2, I-00185 Roma, Italy}
\affiliation{$^{13}$Institute for Nuclear Research, 60-letiya Oktyabrya prospekt 7a, 117312 Moscow, Russia}
\affiliation{$^{14}$INFN - Laboratori Nazionali di Frascati, via E. Fermi 40, I-00044 Frascati, Italy}
\affiliation{$^{15}$Centro Siciliano di Fisica Nucleare e Struttura della Materia, viale Andrea Doria 6, I-95125 Catania, Italy}

\date{\today}

\begin{abstract}
We report on new measurements of the beam asymmetry for $\omega$ photo-production on proton and neutron
in Hydrogen and Deuterium targets from the GRAAL collaboration.
The beam asymmetry values are extracted from the reaction threshold
($E_{\gamma}$ = 1.1 GeV in the free nucleon kinematics)
up to 1.5 GeV of incoming photon energy.
For the first time both the radiative and the three-pion decay channels are simultaneously investigated on the free proton.
Results from the two decay channels are in agreement and
provide important constraints for the determination of resonant state contributions
to the $\omega$ production mechanism.
First experimental results on the deuteron allow the extraction of the $\Sigma$ beam asymmetry on quasi-free nucleons.
Comparison of the results for free and quasi-free kinematics on the proton
shows a generally reasonable agreement,
similar to the findings in pseudo-scalar meson photo-production reactions.
For the first time measurements on quasi-free neutrons are available, showing that both the strength and the angular distributions
of the beam asymmetry are sensibly different from the results on the proton target.
\end{abstract}



\maketitle
%
%
\section{INTRODUCTION}
\label{intro}
The observation of nucleon excited states is fundamental for the unraveling of its internal structure.
It is now understood that dynamic chiral symmetry breaking is responsible for dressing the QCD current-quark masses,
which evolve into constituent-quarks as their momenta decrease \cite{Rob_03}.
This creates a theoretical justification for cons\-tituent-quark models which have been used
to predict  the nucleon spectrum \cite{Cap_94,Riska_01}.
However, important differences are still observed between the experimental nucleon spectrum and the theoretical ones;
several predicted excited states have not been observed yet and have been called {\itshape missing resonances}.

Since the nucleon excited states decay strongly with meson emission,
meson production experiments on the nucleon are the ideal environment to search for missing resonances.
Most of our knowledge of the resonance properties comes from pion induced reactions.
Koniuk and Isgur \cite{KoIs_80} interpreted the missing resonances as states which are weakly coupled to the pion channels
and suggested to access them using an electromagnetic probe
and detect final states such as $\eta N$, $\omega N$ or $K \Lambda$.

The study of $\omega$ photo-production on the nucleon is interesting for several reasons.
The reaction threshold lies in the third resonance region, providing access to higher mass resonances.
Because of the $\omega$ meson relatively small decay width ($\Gamma$=8 MeV),
its experimental invariant mass spectrum is peaked and relatively easy to distinguish from the non-resonant background. 
Moreover, isospin conservation allows only the excitation of $N^*(I={1\over 2})$ resonances in the reaction mechanism,
and only a limited number of states is involved in the excitation spectrum. 

The extraction of $N^*$ parameters from photo-producti\-on data would ideally proceed
through experimental determination of reaction amplitudes for all spin and isospin states,
using a model-independent procedure, and then relying on theoretical models
to separate the resonant contributions from the non-resonant processes.
In the case of vector meson photo-production on the nucleon,
twelve independent complex amplitudes are necessary to completely specify the reaction in the spin space \cite{pick_96},
which requires measurements of at least twenty-three independent polarization observables,
including the unpolarized differential cross section.
Since the electromagnetic interaction is not isospin invariant,
isoscalar and isovector amplitudes are mixed with opposite signs in the $\omega$ photo-production on proton and neutron.
Data on both proton and neutron target are therefore necessary to extract all reaction amplitudes from experimental results
and perform a so-called "complete experiment".
As this task remains very difficult for vector meson photo-production experiments,
the extraction of resonance parameters from the data relies mostly on theoretical models,
which are constrained by the inclusion of new polarization observables in the database.

First measurements on $\omega$ meson photo-production date back to the 60's and 70's
\cite{BCG_67,ABBHHM,Eis_69,Ba_73,Cl_77}.
Differential cross sections were measured
at incoming photon energies ranging from threshold to 9.3 GeV.
An exponential decay of the differential cross section with increasing values of -$t$
(square of the four-momentum transfer) was observed at the higher energies.
A similar trend was also observed, at low momentum transfer,
at energies close to the reaction threshold.
This diffractive behavior was interpreted within the vector dominance model (VDM)
as a direct $\gamma$-$\omega$ coupling followed by the elastic scattering of the $\omega$ meson on the proton target. 
More recently \cite{Fr_96} this process has been described in terms of
natural-parity exchange (Pomeron)
and unnatural-parity exchange ($\pi^0$) in the $t$-channel;
the former dominates at high energies
while the latter is the most important contribution at energies close to threshold
and for small $|t|$.
The role of the two contributions may be disentangled by exploiting the additional information
provided by polarization observables such as the spin density matrix elements \cite{Ba_73}.
At large momentum transfer, deviations from the pure diffractive behavior
are associated to $s$- and $u$-channel contributions \cite{Cl_77}, including resonance excitation.
A more direct way to estimate the contribution of intermediate resonant states
consists in measuring polarization observables:
in particular, the beam asymmetry $\Sigma$
is expected to be not null only if $s$- and $u$-channel terms are involved in the
$\omega$ photo-production process. 

In the past decade the study of baryon resonances has motivated several facilities
to measure the $\omega$ meson differential cross section on the proton, with very high precision,
in the energy region from threshold to 2.8 GeV \cite{Saphir_98,Saphir_03,Clas_03,Eid_06,Clas_09}.
Several theoretical interpretations attempted the extraction of resonant contributions,
reaching conflicting results \cite{Zhao_98,Zhao_01,Oh_01,Ti_02,Gie_05,BG_05,Paris_09,EBAC_07},
but all agreeing that beam polarization observables are needed to constrain the dynamics of the $\omega$ photo-production reaction.

At present, beam polarization asymmetry has been measured only on the proton target
from energy threshold up to 1.5 GeV
by some members of the GRAAL collaboration \cite{Eid_06},
studying the $\omega \rightarrow \pi^+ \pi^- \pi^0$ decay channel,
and up to 1.7 GeV by the CBELSA/TAPS collaboration \cite{CBELSA},
investigating the $\omega$ radiative decay channel.
The two results are not in agreement within the quoted errors and
show different angular distributions.

We are report ing on new high precision results of beam asymmetry for
$\omega$ photo-production off proton.
For the first time, the $\omega$ meson is investigated both
in the $\omega \rightarrow \pi^+ \pi^- \pi^0$ decay 
and in the $\omega \rightarrow \pi^0 \gamma$ radiative decay, simultaneously.
We also provide the very first results of the beam asymmetry
for $\omega$ photo-production off neutron,
measured in the quasi-free kinematics on a deuteron target.

%
%
\section{THE GRAAL EXPERIMENT}
\label{sec:GRAAL}
The GRAAL experiment was located at the European Synchrotron Radiation Facility (ESRF) in Grenoble (Fran\-ce),
where it has been taking data from 1995 to 2008.

A linearly polarized photon beam impinged on a liquid $H_2$ or $D_2$ target,
and the final products were detected by the large solid angle detector LAGRAN$\gamma$E
(Large Acceptance GRaal-beam Apparatus for Nuclear $\gamma$ Experiments).
The photon beam was produced by the Compton backscattering
of low-energy polarized photons from an Argon laser,
against the 6.03 GeV electrons circulating inside the ESRF storage ring
(see \cite{LADON_96} for more details on backscattered photon beams).

The UV laser line was used to produce a backscattered photon beam,
covering the energy range up to 1.5~GeV.
By the use of the far-UV laser line,
the investigated energy range was extended up to 1.55 GeV for the reaction off free proton.
A tagging system, located inside the electron ring, provided an event-by-event measurement of the photon beam energy,
with a resolution of 16 MeV (FWHM).

Since the electron involved in the Compton scattering is ultra-relativistic,
its helicity is conserved in the process at backward angles
and the outgoing photon retains the polarization of the incoming laser beam ($\simeq$ 100\%).
The correlation between photon energy and polarization is calculated with QED \cite{Babusci_95}
and is higher than 68\% in the energy range from the reaction threshold to 1.55 GeV.
During the data taking a half-wavelength plate was used to rotate the beam polarization by $90^{\circ}$
(vertical to horizontal and back) at intervals of about twenty minutes,
in order to collect data in the same experimental conditions with both polarization directions.

The LAGRAN$\gamma$E detector can be divided into two angular regions:

1. the central region ($25^{\circ} \leq \theta \leq 155^{\circ}$ in the laboratory frame) consisting of:
\begin{itemize}
  \item two cylindrical  multi-wire proportional chambers,
  used for charged particle tracking, having an angular resolution of $3.5^{\circ}$ and $4^{\circ}$
  for polar and azimuthal angles, respectively \cite{Annick_07};
  \item an inner plastic scintillator barrel, used for the discrimination between charged and neutral particles;
  \item the BGO electromagnetic calorimeter (see \cite{PLS_96}, \cite{Castoldi_98} and \cite{Ghio_98}),
  made of 480 crystals and optimized for photon detection
  with an energy resolution of 3\% at 1 GeV and
  angular resolution of $6^{\circ}$ and $7^{\circ}$ (FWHM)
  for the polar and azimuthal angles, respectively;
  it has good performances also for proton detection for kinetic energies up to 400 MeV;
\end{itemize}

2. the forward region ($\theta \leq 25^{\circ}$ in the laboratory frame) consisting of:
\begin{itemize}
  \item two planar multi-wire proportional chambers, for char\-ged particle tracking with a resolution of
  $\simeq 1.5^{\circ}$ and $\simeq 2^{\circ}$ (FWHM) for polar and azimuthal angles, respectively;
  \item a double wall of plastic scintillator bars,
  with a time resolution of 300 ps,
  for time of flight (TOF) and
  impact coordinates measurement of charged particles.
  It may be used for proton/pion discrimination and
  for the precise proton energy calculation from TOF measurement;
  \item a shower wall,
  with a time resolution of 600 ps  \cite{Slava_02},
  for TOF and impact coordinates measurement
  for both charged and neutral particles.
  It may be used for neutron/photon discrimination and for
  neutron energy calculation from TOF measurement.
  Neutral particle direction of flight may be determined with a resolution
  of $\simeq$ $3^{\circ}$ (FWHM), for both polar and azimuthal angles.
\end{itemize}

At the end of the beamline, two flux monitoring detectors were used.
The first one, with respect to the beam direction, was composed of two plastic scintillators
preceded by an Aluminum foil to convert photons into electron-positron pairs,
while a third plastic scintillator before the Aluminum foil was used as a veto for the upstream background.
Its detection efficiency was low ($\simeq3\%$) to avoid pile-up effects during data taking.
The second flux monitor consisted of a uniform array of plastic scintillating fibers and lead \cite{Bellini_97}
(spaghetti calorimeter).
Its photon detection efficiency approaches to 1
and it was used to calibrate the efficiency of the former monitor,
using the low intensity Bremsstrahlung beam.
A detailed description of the LAGRAN$\gamma$E apparatus can be found in \cite{Bartalini_06}
and a schematic representation of the whole GRAAL set-up is shown in fig.1 of the same reference.

%
%
\section{DATA ANALYSIS}
\label{sec:analysis}
The three-pion decay $\omega \rightarrow \pi^+ \pi^- \pi^0$ (B.R. $\simeq$ 89\%) and the radiative decay
$\omega \rightarrow \pi^0 \gamma$ (B.R. $\simeq$8.3\%) are the main
decay channels of the $\omega$ meson \cite{10_partbooklet}.
The exclusive measurement of the $\omega$ meson photo-production on the nucleon
in the radiative decay channel requires the detection of a nucleon and
three photons in the final state.
If the angles and kinetic energies of all particles in the final state are measured,
the kinematics of the reaction is over-determined.
Since the GRAAL apparatus is optimized for photon detection and no other competing reaction has the same final state,
we expect that events from this channel can be selected and separated from background.
The three-pion decay channel requires the detection of two charged pions
in addition to a $\pi^0$ and a nucleon. 
It is important to point out that the GRAAL detector does not provide information on charged pion energies,
and that these must be deduced from kinematics constraints.
In the data analysis of the free proton ($H_2$ target),
the target nucleon can be considered at rest and
momentum conservation relations can be used
to calculate the momentum strength for each of the charged pions.
However, in the investigation of $D_2$ data in the quasi-free kinematics,
it becomes impossible to estimate the energy (or the momentum) of the charged pions in the final state,
due to the unknown Fermi momentum of the target nucleon.
Moreover, the direct $\pi^+ \pi^- \pi^0$ photo-production cannot be separated from the $\omega$ photo-production reaction,
and it must be properly evaluated and subtracted.

Due to these considerations, the radiative and the three-pion decays have simultaneously been investigated
for the free proton target only.
Since the beam asymmetry measurement is independent from the $\omega$ meson decay mode,
assuming that the events are integrated over the whole decay phase space,
a comparison of the results obtained for the two decay channels provides a strong check on systematic errors
and on their stability when different event selection techniques are used.
The data analysis procedure applied for the radiative decay on the free proton is then extended to the deuteron data set,
in order to study the $\omega$ photo-production on both quasi-free proton and quasi-free neutron targets.

The first step in the analysis, common to all data sets and reaction channels,
consists in the association of all signals from the LAGRAN$\gamma$E detector to the particle tracks.
In the central region,
those tracks where at least one detector has been found in geometrical and time coincidence with the BGO calorimeter,
are interpreted as produced by charged particles,
while neutral particles are expected to leave a signal in the BGO calorimeter only.
In the forward region,
those tracks where at least two detectors provided a signal
(either the two layers of the scintillating wall or the wire chambers and first layer of the scintillating wall)
are interpreted as produced by charged particles,
while neutral particles are associated to signals observed in the shower wall only.

A primary selection criterion in the analysis of all the reaction channels is
the choice of a proper number of charged and neutral particle tracks in the apparatus
as explained in detail in the following subsections.

%
%
\subsection{$\omega$ photo-production on the free proton: the radiative decay channel
($\omega \rightarrow \pi^0 \gamma$)}
\label{subsec:fp_raddecay}
The analysis is performed for all the events showing at least one charged track and three neutral signals,
among which at least two neutral particles are
detected by the BGO calorimeter.

The next step consists in associating the tracks to the final state particles of the
$\gamma p \rightarrow \omega p \rightarrow \pi^0 \gamma p \rightarrow \gamma \gamma \gamma p$
reaction, without discarding events.
For each charged track the missing mass of the reaction
$\gamma p \rightarrow p' X$
($M_{miss} = \sqrt{(\tilde{p}_{\gamma}+\tilde{p}_p-\tilde{p}_{p'})^2}$)
is calculated, using the measured tagged photon energy.
The charged track whose missing mass is the closest to the $\omega$ mass value
($M_{\omega}$ = 782.57 MeV \cite{10_partbooklet})
is interpreted as the final-state proton. 
The three neutral signals whose invariant mass is the closest to the $\omega$ mass value
are interpreted as the three final-state photons.
If one of the photons is
detected in the forward direction,
its energy is calculated by imposing the energy balance of the reaction.
Among the three selected final-state photons, the couple whose invariant mass is the closest to the $\pi^0$ mass value
is matched to the $\pi^0 \rightarrow \gamma \gamma$ decay.
The energy and angles of the $\omega$ meson
(and of the $\pi^0$ from its decay)
are calculated using the momentum
of the three final-state photons.

Among all measured quantities, the incoming photon energy and the outgoing proton angles
are the ones measured with the best resolution. 
Starting from these quantities and imposing two-body four-momentum conservation
of the $\gamma p \rightarrow \omega p$ reaction,
the proton energy and the $\omega$ meson momentum are calculated, obtaining a new set of variables
($E_{\omega}^{calc}$,$\theta_{\omega}^{calc}$, $\phi_{\omega}^{calc}$ and $E_P^{calc}$)
which can be used in the data selection procedure.

\begin{figure}[htbp]
  \begin{center}
  \resizebox{0.5\textwidth}{!}{\includegraphics{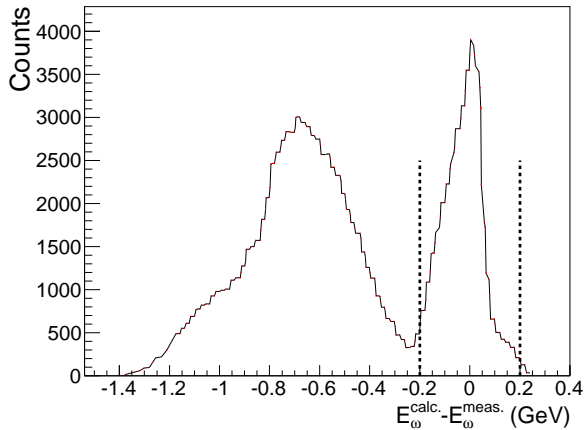}}
  \caption{Experimental distribution of the difference between
  the calculated and the measured energy of the $\omega$ meson.
  The two vertical lines indicate the range of the selected energy region (first selection cut).
  Simulation studies show that the background events mainly populate the bump structure
  corresponding to the largest difference between calculated and measured energy.}
  \label{fig:energy}
  \end{center}
\end{figure}

\begin{figure}[htbp]
  \begin{center}
  \resizebox{0.5\textwidth}{!}{\includegraphics{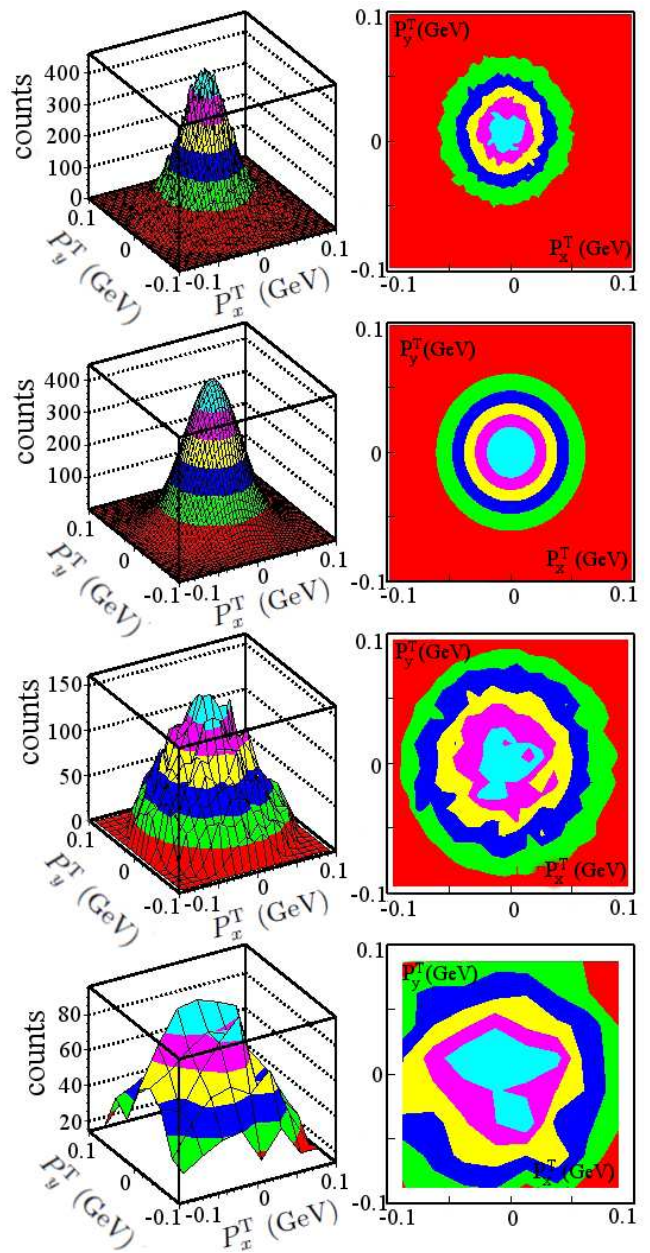}}
  \caption{Transverse momentum distribution of final-state particles:
  $x$ $vs$. $y$ distribution in a 3-D image (left panel)
  and corresponding level curves (right panel).
  First and second rows (section III.A): experimental distribution and bi-gaussian fit
  on the free proton target.
  Third and fourth rows (section III.C): experimental distribution
  on the quasi-free proton and on the quasi-free neutron, respectively.}
  \label{fig:ptrans}
  \end{center}
\end{figure}

On the basis of simulation studies,
it is justified to clean the data set selecting the sole events which do not show
any other signals from the detector, except the ones already interpreted as the final-state proton or photons.
This selection is performed requiring that no other signal appears in the forward detector,
and accepting events for which an energy of max. 5 MeV is deposited
in the BGO calorimeter by extra-signals.
These conditions reject background events but still save
$\omega$ events for which low-energy particles are produced
by the passage of the nucleon in the BGO crystals
or by albedo of the detected photons.
\\
The two remaining major background sources are identified as events from the
\begin{equation}\label{eqn:one}
 \gamma p \rightarrow \pi^0 \pi^0 p \rightarrow \gamma \gamma \gamma \gamma p
 \end{equation}
reaction channel, when one photon is undetected,
and from the
\begin{equation}\label{eqn:two}
 \gamma p \rightarrow \pi^0 p \rightarrow \gamma \gamma p
 \end{equation}
 reaction channel,
when a third random neutral signal in the detector is erroneously interpreted as the third final-state photon.
Selection criteria are determined by analyzing 9$\times$10$^6$ simulated events
from all possible photo-reaction channels on the proton target,
in the incoming photon energy range covered by the experiment.
The aim is to maximize the event selection efficiency and minimize the background contamination.
The following selection criteria are identified:
\begin{enumerate}
  \item the absolute value of the difference between the measured and calculated energy of the $\omega$ meson
  must be smaller than 200 MeV
  ($|E_{\omega}-E_{\omega}^{calc}| <$ 200 MeV) (see fig.\ref{fig:energy},
  where background events are concentrated in the first bump);
  \item since four-momentum conservation implies that the total
 transverse momentum of the detected particles must be
 null, the experimental correlation of the transverse
 momentum components of the reaction
 $P_x^{TOT}$ $vs$. $P_y^{TOT}$ is calculated from the proton
 and $\omega$ meson tri-momentum components and
 can be fitted using a bi-gaussian distribution:
  $$\frac{(P_x^{TOT})^2}{\sigma_x^2}+\frac{(P_y^{TOT})^2}{\sigma_y^2} \le n^2$$
  where $\sigma_x = \sigma_y$ = 30 MeV/c are the fitted widths
  associated to the experimental distributions of the variables $P_x^{TOT}$ and $P_y^{TOT}$, respectively
  (fig.\ref{fig:ptrans}, upper panels).
  The event selection is performed choosing $n$=3;
  this cut embeds the constraint on the co-planarity of the final-state particles;
  \item the proton missing mass must be at least 200 MeV;
  \item the invariant mass of the three final-state photons must be at least 680 MeV.
\end{enumerate}


\begin{figure}[htbp]
  \begin{center}
  \resizebox{0.5\textwidth}{!}{\includegraphics{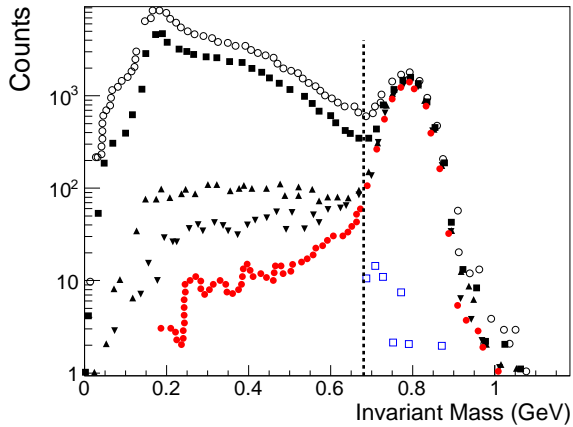}}
  \caption{Effects of the selection criteria on the $\omega$ mass distribution for simulated data.
  Empty black circles: all events;
  full black squares: distribution after the cut on the number of residual signals in the forward detector;
  full black upward triangles: distribution after the cut on the energy deposited by residual signals in the BGO calorimeter
  (required to be smaller than 5 MeV);
  full black downward triangles: distribution after the kinematical cut on the difference between the calculated
  and the measured energy of the $\omega$ meson (cut number 1)
  and the transverse momentum conservation (cut number 2);
  full red circles: distribution after the cut on the proton missing mass (cut number 3).
  The black vertical line corresponds to the condition applied on the three final-state photons
  invariant mass (cut number 4).
  The distribution of the remaining background events is shown by the empty blue squares.
  Background contamination is estimated to be smaller than 6\% of all selected events from simulation studies.
  }
  \label{fig:ome_mass}
  \end{center}
\end{figure}

First and second cuts, shown in fig.\ref{fig:energy} and fig.\ref{fig:ptrans} respectively,
are very effective in suppressing the two-$\pi^0$ photo-production background;
the third cut is needed to suppress single $\pi^0$ photo-production events.
The effect of the event selection constraints is monitored using the $\omega$ meson invariant mass spectrum,
as shown in fig.\ref{fig:ome_mass}.
The residual background event contamination,
shown by the empty blue squared points in fig.\ref{fig:ome_mass},
is estimated through the simulation to be less than 6\% of all selected events.

%
%
\subsection{$\omega$ photo-production on the free proton: the 3-pion decay channel
($\omega \rightarrow \pi^+ \pi^- \pi^0$)}
\label{subsec:fp_pionsdecay}
The analysis is performed for all events with at least three charged tracks and two neutral signals.
The charged track corresponding to the final-state proton is identified
according to the same missing mass criterion that is applied for the radiative decay (see previous section).
For the $\omega$ meson reconstruction,
all possible combinations of two neutral signals in the BGO calorimeter
and two charged tracks in the whole detector are considered,
with unknown energies and charge signs for the two candidate charged pions.
Among all neutral signal pairs detected in the BGO calorimeter,
the one whose invariant mass is closest to the $\pi^0$ mass value is selected
and identified as the couple of photons from the $\pi^0$ decay.
The selected photons are then combined with all the residual charged track couples,
considered as candidate charged pions from the $\omega$ decay.
As already pointed out, the energy of charged pions is not measured in the detector.
Nevertheless, the charged particle track angles are known,
and transverse momentum conservation law is reduced to two
equations, which allow the extraction of the two unknown pion momenta.
The pair which better satisfies the longitudinal momentum conservation
is selected and matched to the charged pions from the $\omega$ decay.\\
In order to obtain a better coverage of the phase space,
the possibility that one photon from the $\pi^0$ decay is emitted in the forward direction is also investigated.
The combinations of one neutral signal from the BGO calorimeter,
one photon in the shower wall and two charged tracks in the whole apparatus are considered.
In this case not only the charged pion energies, but also the energy of the forward photon is unknown.
The three-momentum conservation law is used to determine the charged tracks momenta
and the forward photon energy.
When more than two neutral signals appear as candidate final-state photons,
the pair whose ($\gamma_1$,$\gamma_2$) invariant mass is closest to the $\pi^0$ mass is selected.
All kinematical variables of the $\omega$ meson
($E_{\omega}$, $\theta_{\omega}$ and $\phi_{\omega}$)
and of the $\pi^0$ meson
($E_{\pi^0}$, $\theta_{\pi^0}$ and $\phi_{\pi^0}$) can then be calculated.
Finally, a two-body kinematics reconstruction of the
$\gamma p \rightarrow \omega p$ reaction is attempted,
as done previously for the radiative decay channel.

The most important difference, 
between the radiative decay analysis and the current one,
lies in the event selection procedure.
In the case of the three-pion decay,
background events arise mainly from the direct three-pion reaction:
$$\gamma p \rightarrow \pi^+ \pi^- \pi^0 p.$$
This represents the physical background for our process and there are no cuts allowing for background rejection,
since both $\omega$ photo-production and background events satisfy the same kinematics.
Therefore, a background subtraction technique must be applied.
Two cuts are previously used to clean the data set:
the three-pion invariant mass must be smaller than 2 GeV
and the invariant mass of the two photons must lay in the range from 100 MeV up to 170 MeV.
Then a fitting procedure for the estimation of signal and background events is developed.

\begin{figure}[htbp]
  \begin{center}
  \resizebox{0.5\textwidth}{!}{\includegraphics{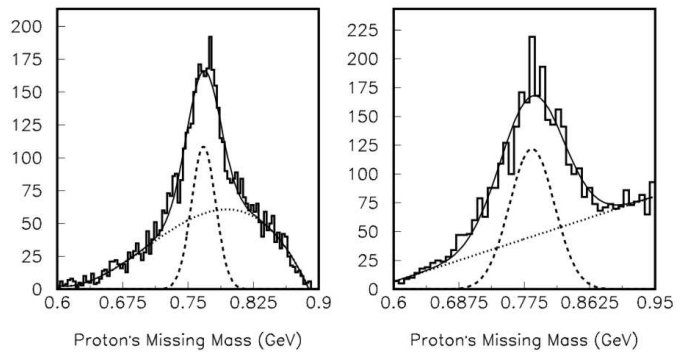}}
  \caption{Examples of the fit on the proton missing mass distribution.
  Dashed line: signal contribution.
  Dotted line: back-ground contribution.
  Solid line: global fit.
  In the left panel, experimental distribution and corresponding fit
  for $E_{\gamma}$=1.26 GeV, $\theta_{\omega}^*=18^{\circ}$, $\phi_{\omega}=225^{\circ}$ for the vertical polarization.
  The $\omega$ event distribution is described by a gaussian function
  while the background event distribution, mainly due to the non resonant three-pion production,
  is described by a 3$^{rd}$ order polynomial function.  
  In the right panel, observed distribution and corresponding fit
  for the highest energy bin $E_{\gamma}$=1.48 GeV,
  $\theta_{\omega}^*=18^{\circ}$, $\phi_{\omega}=15^{\circ}$ for the vertical polarization.
  At this energy, only the low energy tail of the background distribution is visible
  under the gaussian peak and it is described by a 1$^{st}$ order polynomial function.}
  \label{fig:fit}
  \end{center}
\end{figure}

Two distributions can be used to extract the number of events:
the final-state proton missing mass 
($\gamma p \rightarrow p' X$)
and the three-pion invariant mass distributions.
The first one is more reliable, since it is directly calculated from measured variables (proton energy and angles).
The fit of the proton missing mass distribution is performed at fixed values of the incoming photon energy
$E_{\gamma}$ (4 bins), of the $\omega$ polar and azimuthal angles
($\theta_{\omega}^*$ and $\phi_{\omega}$)
in the center-of-mass frame (5 and 16 bins, respectively)
and for each polarization state (vertical and horizontal).
A total of 640 bins is fitted.
As suggested by simulation studies,
the fitting function is the sum of a gaussian distribution and of a 3$^{rd}$ order polynomial function,
which reproduce the signal and the background respectively (fig. \ref{fig:fit}, left panel).
In the last energy bin only the low energy tail of the background distribution is visible 
under the gaussian peak, and it can be approximated by a 1$^{st}$ order polynomial function
(fig. \ref{fig:fit}, right panel), both for experimental and for simulated data.\\
The final $\omega$ photo-production event number for each bin is estimated in two different ways:
i) as the integral of the gaussian function obtained for the signal fit;
ii) as the difference between the total number of histogram events
subtracted by the events integrated from the background fit.
The results from the two techniques are in agreement within the errors.

%
%
\subsection{$\omega$ photo-production on the quasi-free nucleon: the radiative decay channel
($\omega \rightarrow \pi^0 \gamma$)}
\label{subsubsec:qf}
Data analysis is performed both on the proton and on the neutron target in the participant/spectator description.
The reaction is thought to take place on one of the nucleons (the participant), which has a Fermi motion,
while the second one (the spectator) is not involved in the process.
The event selection procedure, optimized for the proton target and the $\omega$ radiative decay channel,
may be extended to the deuteron data-set.
The final-state nucleon is identified according to the missing mass criterion.
Since knowledge of the nucleon energy is necessary for the missing mass calculation,
in the case of $\omega$ photo-production on neutron target,
only events with the neutron emitted in the forward direction are analyzed.
The identification procedure of the three final-state photons is identical to the one performed for the free proton target.

Small changes in the analysis appear only in the selection cuts,
which are optimized by analyzing 12$\times$10$^6$ simulated events
from all possible reaction channels on quasi-free nucleons from a deuteron target.
Because of smearing due to the Fermi momentum,
the transverse momentum conservation constraint is applied with a tolerance of 4$\sigma$ (instead of 3$\sigma$)
with $\sigma_x = \sigma_y$ = 70 MeV/c, (see fig. \ref{fig:ptrans} third and fourth rows).
An additional bi-dimensional selection is imposed on the correlation between the three-photon invariant mass
($M^{inv}_{\gamma \gamma \gamma}$)
and the detected nucleon missing mass ($M^{miss}_N$) values:
$$\left( \frac{M_{\gamma \gamma \gamma}^{inv} - M_{\omega}}{\sigma_{\gamma \gamma \gamma}^{inv}} \right)^2
+\left( \frac{M_N^{miss} - M_{\omega}}{\sigma_N^{miss}} \right)^2<n^2$$
where $M_{\omega}$ = 782.57 MeV,
$\sigma_{\gamma \gamma \gamma}^{inv}$ = 60 MeV/c$^2$,
$\sigma_N^{miss}$ = 80 MeV/c$^2$ and n = 3.
In the case of $\omega$ photo-production on the quasi-free proton,
background events come not only from  (\ref{eqn:one}) and (\ref{eqn:two}), but also from
$$\gamma n (p) \rightarrow \rho^- p (p) \rightarrow \pi^0 \pi^- p (p)$$
and
$$\gamma n (p) \rightarrow \pi^0 \pi^- p (p)$$
reactions,
where the $\pi^-$ is erroneously interpreted as a photon.
These events are rejected by requiring that no signal be recorded in the plastic scintillator barrel
in coincidence with a BGO cluster
(not even the ones below hardware threshold in time with the event, which are included in the free-proton analysis).

\begin{figure}[htbp]
  \begin{center}
   \resizebox{0.5\textwidth}{!}{\includegraphics{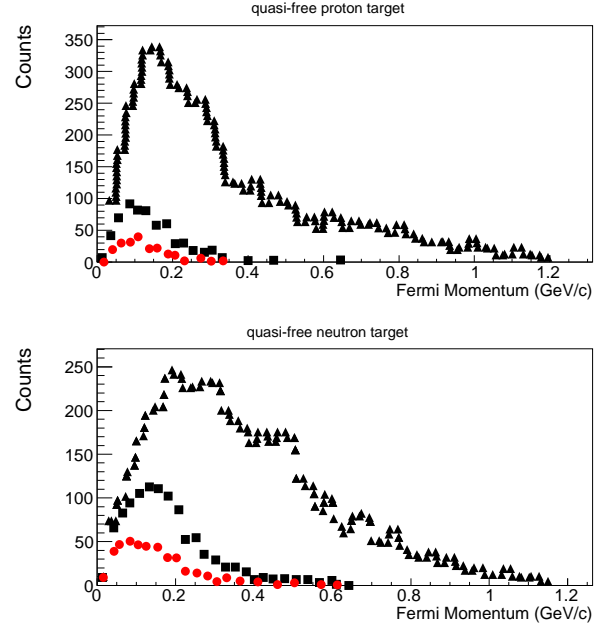}}
   \caption{Effects of events selection criteria on the Fermi momentum distribution for the quasi-free proton (top panel)
  and the quasi-free neutron (bottom panel).
  Fermi momentum is evaluated as described in the text.
  Full black triangles: Fermi momentum distributions after the cut on the transverse momentum;
  full black squares: distributions after the cut on the difference between the measured and the calculated energy
  of the $\omega$ meson,
  bi-dimensional selection on the three final-state photons invariant mass and recoil nucleon missing mass;
  full red circles: Fermi momentum distributions of selected events.}
  \label{fig:pfermi}
  \end{center}
\end{figure}

In the case of $\omega$ photo-production on the quasi-free neutron,
background events come from the
$$\gamma n (p) \rightarrow \pi^0 n (p)$$
reaction if one photon is detected in the forward direction.
In accordance with simulation studies,
these background events appear mainly when the $\pi^0$ is emitted at backward angles and a soft photon reaches the forward region.
For background events, $\pi^0$ and neutron must satisfy the coplanarity condition.
Which is not the case for a $\pi^0$ coming from the $\omega$ decay.
Simulation studies of the difference $\Delta\Phi$ between
the azimuthal angle of the $\pi^0$ ($\phi_{\pi^0}$)
and of the neutron ($\phi_n$)
revealed that for background events $\Delta\Phi$ ranges between 150$^{\circ}$ and 210$^{\circ}$.
In this way, events from $\pi^0$ photo-production off neutron are identified and finally rejected.\\
The residual background events, after the described selection criteria are applied,
are lower than 13\% and 8\% for the quasi-free proton and quasi-free neutron target, respectively.
In fig. \ref{fig:pfermi}, we show the effect of the applied selection criteria on the Fermi momentum distribution,
calculated from the total momentum components of the final state $\vec{p}_{TOT}$ according to:
$$p_F = \sqrt{(p_x^{TOT})^2 + (p_y^{TOT})^2 + (p_z^{TOT} - E_{\gamma})^2} $$
where $E_{\gamma}$ is the incoming photon energy.
Final selected events show a maximum Fermi momentum value of about 400 MeV/c.

%
%
\subsection{Extraction of the $\Sigma$ beam asymmetry}
\label{subsec:sigma}
\begin{figure}[htbp]
 \begin{center}
   \resizebox{0.45\textwidth}{!}{\includegraphics{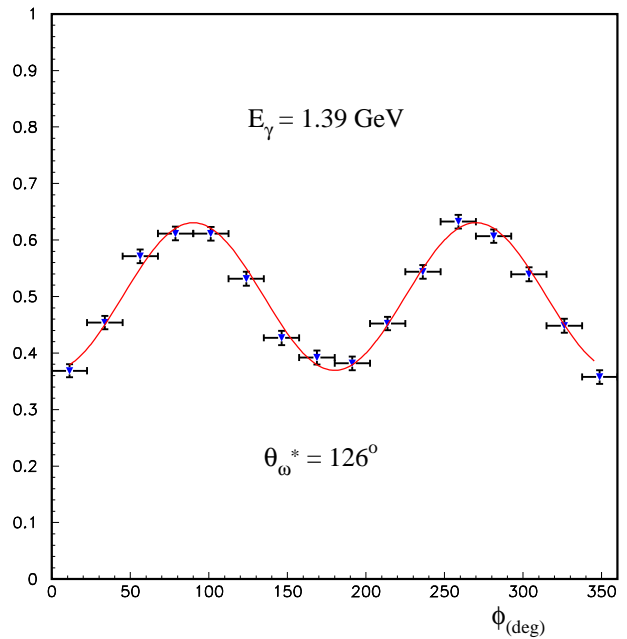}}
 \caption{An example of the azimuthal distribution used for the extraction of the $\Sigma$ beam asymmetry at fixed values of $E_{\gamma}$ and $\theta_{\omega}^*$
 (in this example: $E_{\gamma}$=1.39 GeV and $\theta_{\omega}^*=126^{\circ}$).}
 \label{fig:psigma}
 \end{center}
\end{figure}

The differential cross section of the $\omega$ photo-production reaction with polarized photons
can be expressed in terms of the unpolarized cross section as:
$$\left( \frac{d\sigma}{d\Omega} \right)_{pol} = 
\left( \frac{d\sigma}{d\Omega} \right)_{unp} \{1-P(E_{\gamma})\Sigma(E_{\gamma},\theta_{\omega}^*)\cos 2\varphi \}$$
where $\varphi$ is the difference between the azimuthal angle of the reaction plane $\phi$
and the incident photon polarization vector $\varphi_{\gamma}$
($\varphi = \phi - \varphi_{\gamma}$).
We chose to define the laboratory frame having the $\hat{z}$ component along the photon beam direction,
the $\hat{y}$ component along the vertical direction and the $\hat{x}$ component such that
$\hat{z} = \hat{x} \times \hat{y}$.
If the photon beam is polarized in the horizontal direction ($\varphi_{\gamma}$ = 0) then $\varphi$ = $\phi$;
if the beam is polarized in the vertical direction ($\varphi_{\gamma}$ = $\pi /2$) then
$\varphi$ = $\phi - \pi/2$.
In this reference frame, the polarized cross section can then be expressed
as a function of the reaction plane azimuthal angle $\phi$ as follows:
$$\left( \frac{d\sigma}{d\Omega} \right)_{H,V}=
\left( \frac{d\sigma}{d\Omega} \right)_{unp} \{1 \mp P(E_{\gamma})\Sigma(E_{\gamma},\theta_{\omega}^*) \cos 2\phi \}.$$
From the experimental point of view we can express the number $N_{H,V}(E_{\gamma},\theta_{\omega}^*,\phi)$ of experimental yields, normalized by the flux of incident photons,
for a given polarization as follows:
$$ \frac{N_{H,V}(E_{\gamma},\theta_{\omega}^*,\phi)}{F_{H,V}(E_{\gamma})} =\: d\sigma_{H,V}(E_{\gamma},\theta_{\omega}^*)\: \Delta\Omega \: \varepsilon(E_{\gamma},\theta_{\omega}^*,\phi) \: N_{sc}$$
where: $F_{H,V}(E_{\gamma})$ is the incident photon flux for horizontal/vertical polarization;
$d\sigma_{H,V}(E_{\gamma},\theta_{\omega}^*)\: \Delta\Omega$ is the probability of the reaction in the $\Delta \Omega$ solid angle;
$\varepsilon(E_{\gamma},\theta_{\omega}^*,\phi)$ is the detection and event reconstruction efficiency, which is identical for the two polarization states;
$N_{sc}$ is the number of scattering centers.
The $\Sigma$ beam asymmetry was extracted
at fixed values of $E_{\gamma}$ and $\theta_{\omega}^*$ from the azimuthal distribution of the following ratio:
\begin{equation}\label{eqn:psigma}
  \frac{N_V / F_V}{N_V / F_V+N_H / F_H}=\frac{1}{2}\{ 1+ P\Sigma \cos 2\phi \}
\end{equation}
and its value is not affected by systematic errors on the determination of the efficiency $\varepsilon$.
In fig. \ref{fig:psigma}, an example of this azimuthal distribution, divided into sixteen angular bins, is illustrated.

%
%
\section{RESULTS AND DISCUSSION}
\label{sec:results}
%
%
\subsection{$\omega$ photo-production on the free proton}
\label{subsec:res_fp}
\begin{figure*}
 \centering
 \vspace{2cm}
   \resizebox{0.7\textwidth}{!}{\includegraphics{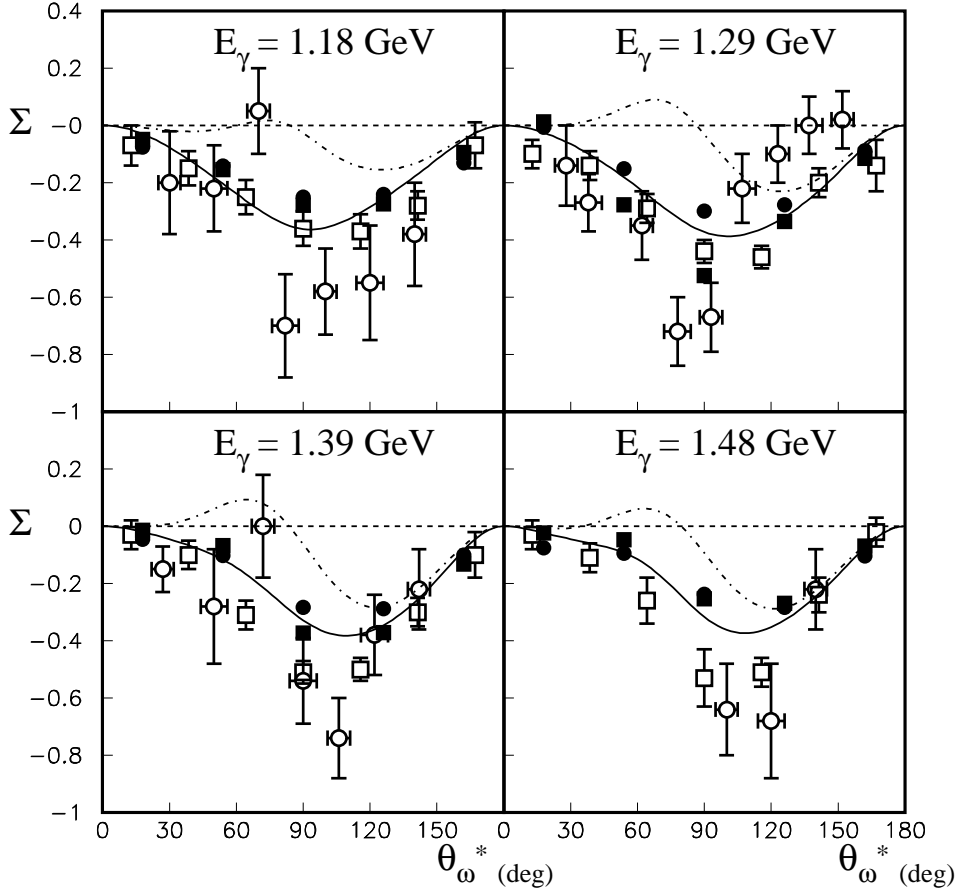}}
 \caption{$\Sigma$ beam asymmetry for the reaction $\vec{\gamma} p \rightarrow \omega p$:
 both the radiative decay (full circles) and the 3-pion decay (full squares) of the $\omega$ meson are investigated.
 Statistical errors are within the marker size.
 Data are compared with previously published results on the three-pion decay channel \cite{Eid_06} (open squares)
 and radiative decay \cite{CBELSA} (open circles).
 Theoretical curves are from the model by Zhao \cite{Zhao_98,Zhao_01}:
 solid lines correspond to the full model, dot-dashed lines do not include contributions from the $P_{13}$(1720) resonance,
 dashed lines correspond to the expected beam asymmetry when $s$ and $u$-channel contributions are not taken into account.}
 \label{fig:sigma_proton_data} 
\end{figure*}

$\Sigma$ beam asymmetry values are extracted for the
$\vec{\gamma} p \rightarrow \omega p$ reaction from threshold up to 1.55 GeV photon energy.
The radiative and the three-pion decay channels are simultaneously investigated for the first time.
Results are shown in fig. \ref{fig:sigma_proton_data} in four incoming photon energy bins
as a function of the $\omega$ meson polar angle $\theta_{\omega}^*$
in the center-of-mass reference frame.
The energy value given for each bin,
corresponds to the average incoming photon energy
weighted by the number of corresponding events.
Full circles and full squares correspond to the results from the analysis
of the $\omega$ radiative decay and the three-pion decay channels, respectively.
Numerical values are listed in tab.\ref{tab:results}.
The $\Sigma$ beam asymmetry values are negative.
They are almost symmetrical with respect to $\theta_{\omega}^* = 90^{\circ}$ in the first two energy bins
and tend to be larger in the backward direction at higher energies.
A strong agreement is observed between the results of the two decay channels,
especially for the highest and the lowest energy bins.
Of the 20 asymmetry values, 3 show a larger discrepancy than
the statistical error, but they are compatible within 2 or 3 $\sigma$.
In the second energy bin and for the third $\theta_{\omega}^*$ bin,
the discrepancy between the two decay channel is resolved within 4 $\sigma$.
This comparison provides a strong check on the stability and reliability of the results
since the analyses of the two channels are based on different techniques.
The identification of the final-state proton is the only procedure common to both analyses.
In order to verify that it does not introduce systematic errors,
an alternative procedure is developed
and applied to the analysis of the radiative decay.
Events with more than one charged track were rejected.
Protons are discriminated from charged pions using graphical cuts on the correlation plots between:
\begin{itemize}
  \item the energy released in the scintillating barrel and the energy deposited in the BGO calorimeter
  (dE/dx $vs$. Energy) in the central part of the detector;
  \item the energy released and the time of flight in the scintillating wall (dE/dx $vs$. ToF)
  in the forward part of the detector.
\end{itemize}

Results obtained with the new procedure are in excellent agreement with the ones from the standard analysis
and we could conclude that no systematic errors should be ascribed to the proton identification procedure.

Three additional checks are performed both for the radiative and the three-pion decay results:
\begin{enumerate}
  \item two sets of 2$\times$10$^6$ events each are
  simulated by a dedicated Montecarlo generator
  based on GEANT3 \cite{Geant3} with different input asymmetry values.
  In both cases the extracted beam asymmetry values are in a strong agreement with the input ones,
  indicating that no modification of the beam asymmetry distribution is
  introduced by the data analysis procedure;
  \item beam asymmetry values are
  extracted by varying the $\theta_{\omega}^*$ and $\phi_{\omega}$ binnings.
  The results prove
  stable and independent from the binning choice;
  \item beam asymmetry values are
  extracted also for background events alone, in several binning configurations.
  They are always compatible with zero.
  Therefore, if background events are erroneously included among the selected $\omega$ events,
  they would affect the asymmetry by reducing its magnitude.
  Since the two decay channels of the $\omega$ meson are
  analyzed with totally independent procedures,
  based on event selection for the radiative decay and background subtraction for the three-pion decay,
  and since they are characterized by different background reactions,
  the residual background contributions are
  expected to affect final results differently for the two channels.
  Due to the agreement of the results coming from the two decay channels
  (see fig. \ref{fig:sigma_proton_data}), we can state that background contamination
  is strongly suppressed in both channels.
\end{enumerate}

Additional systematic uncertainties due to non-unifor\-mities of the experimental acceptance,
that modify the $\omega$ final-state particle distribution,
have been investigated by adding a $\sin(2\phi)$ term to the relation (\ref{eqn:psigma}),
which is used to fit the normalized event azimuthal distribution.
No noticeable difference is found for the final results.

In fig. \ref{fig:sigma_proton_data} the comparison with published data is shown:
open squares are from a different analysis of the GRAAL data performed by some members of the collaboration \cite{Eid_06}
for the three-pion decay channel.
While present results show strong agreement within quoted errors,
in the first two energy bins,
larger asymmetry values are obtained by \cite{Eid_06} at higher energies.
The main difference between the two analysis procedures lies in the tracking reconstruction of the charged particles:
this work fully exploits the good angular resolution provided by the multi-wire proportional chambers,
which are not included in the analysis by \cite{Eid_06}.

Open circles shown in fig.~\ref{fig:sigma_proton_data} are from the CB-ELSA/ TAPS collaboration \cite{CBELSA},
for the $\omega$ radiative decay channel.
These angular distributions,
at fixed values of the incoming photon energy $E_{\gamma}$, show very different trends.
The data analysis from \cite{CBELSA} is characterized by much higher background levels compared to this work,
and the technique of background subtraction on the $\pi^0 \gamma$ invariant mass spectrum is used.

\begin{figure*}
  \centering
  \vspace{2cm}
 \resizebox{0.7\textwidth}{!}{\includegraphics{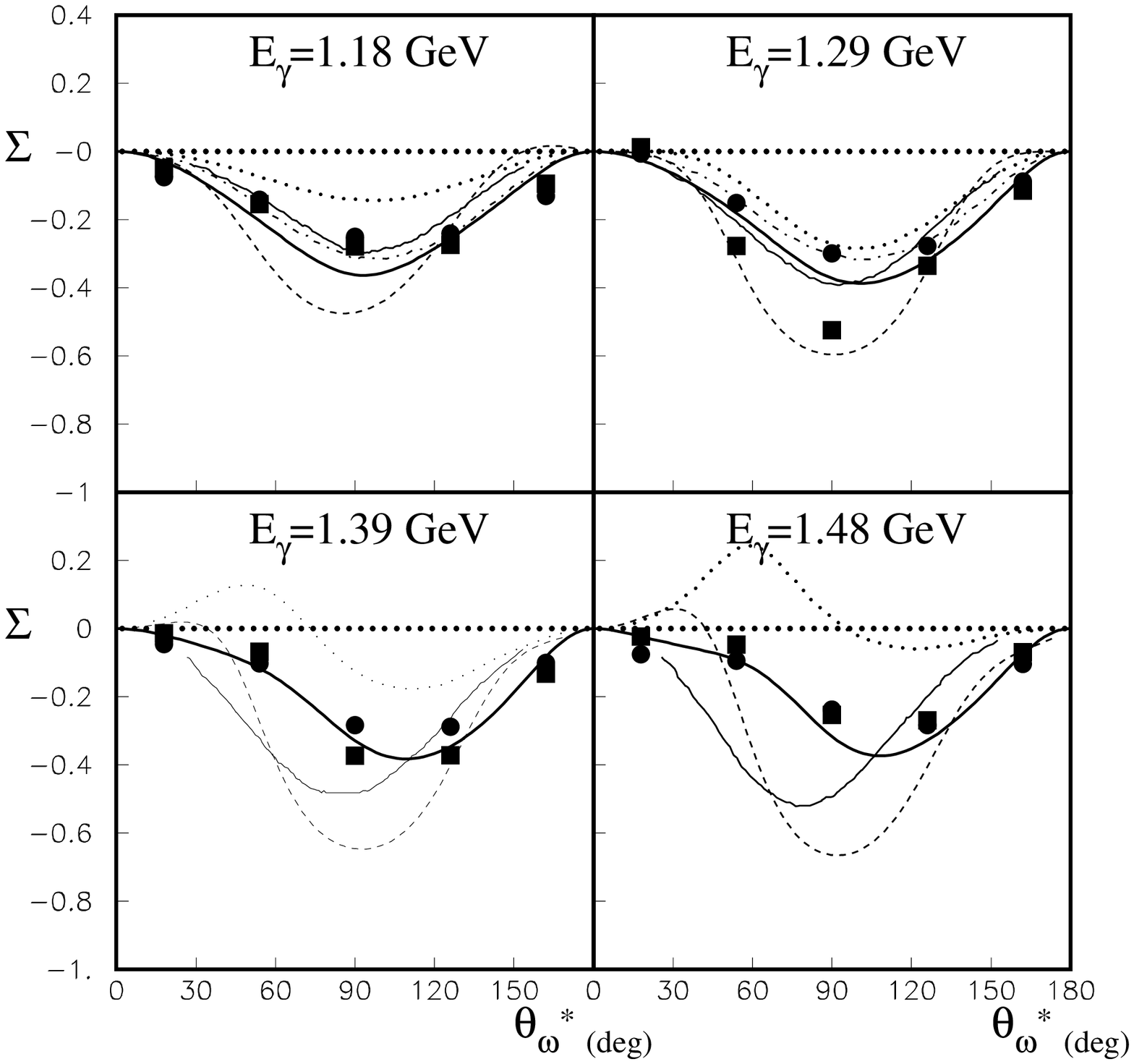}}
 \caption{$\Sigma$ beam asymmetry for the reaction $\vec{\gamma} p \rightarrow \omega p$.
 Full circles are the results of the present analysis for the radiative decay channel;
 full squares are the results of the present analysis for the three-pion decay channel.
 Thin lines are from Bonn-Gatchina PWA \cite{BG_05};
 dot-dashed lines are the predictions from \cite{Ti_02} at threshold;
 dashed lines are the results from \cite{EBAC_07};
 dotted curves are predictions from Giessen model \cite{Gie_05};
 solid curves are from \cite{Zhao_98,Zhao_01}.}
 \label{fig:sigma_proton_theory} 
\end{figure*}

In the last decade, several models were developed
in the attempt of determining the role and the properties of nucleon resonances
from $\pi$N and $\gamma$N reaction data.
The model in \cite{Zhao_98,Zhao_01} uses an effective Lagrangian approach,
based on the SU(6)$\times$O(3) constituent quark model,
with meson-quark couplings adjusted to fit differential cross-section data.
Unnatural parity exchanges ($\pi^0$) and natural parity exchanges (Pomeron) in the $t$-channel are phenomenologically included.
The Moorhouse selection rule \cite{Moorhouse_66} reduces the accessible nucleonic states from the [\textbf{70},$^4$\textbf{8}] representation,
and only eight states from the lowest harmonic oscillator basis contribute to the $\omega$ photo-production reaction
in the $s$-channel. Full predictions for the $\Sigma$ beam asymmetry are shown in fig. \ref{fig:sigma_proton_data}
and fig. \ref{fig:sigma_proton_theory} as solid lines
and are found in generally good agreement with the results of this work.
Dot-dashed lines of fig. \ref{fig:sigma_proton_data} do not include the contribution from the $P_{13}$(1720),
showing a large sensitivity of the polarization observable to this resonance.

An alternative model based on an effective Lagrangian approach is developed in \cite{Oh_01,Ti_02}
including all twelve $N^*$ lowest energy resonances, up to spin J = 7/2,
whose empirical helicity amplitudes of $\gamma N \rightarrow N^*$ transitions are listed by \cite{10_partbooklet}.
Predictions are available only close to the reaction threshold, and are shown in fig. \ref{fig:sigma_proton_theory}
for the two lowest energy bins as dotted-dashed lines.
The model finds that the dominant contribution comes from the excitation of the $F_{15}$(1680) state
and is in fair agreement with our results.

Comparison with the Bonn-Gatchina partial wave analysis \cite{BG_05}
is also shown in fig. \ref{fig:sigma_proton_theory} as continuous thin lines.
The far dominating J = 3/2 wave is associated with the $P_{13}$(1720) resonance,
in accordance with results from \cite{Zhao_01}.
Overestimation of the $\Sigma$-beam asymmetry is obtained at the highest energies,
while agreement is found closer to threshold.

A coupled-channel effective Lagrangian approach, including
$\gamma$N , $\pi$N, 2$\pi$N, $\eta$N and $\omega$N final states,
is presented in \cite{Gie_05},
including all known nucleon states with spin J = 1/2, 3/2 and 5/2 up to masses below 2 GeV.
Results are shown in fig. \ref{fig:sigma_proton_theory} as dotted curves.
They show that the dominating contributions arise from the $D_{13}$ partial wave,
mainly due to the non resonant $\pi^0$ exchange, and from the spin-5/2 resonances
$D_{15}$(1675) and $F_{15}$(1680).
Agreement with our results is visible only for the second energy bin,
while a predicted change of sign at the highest energies is not observed in our results.

Another prediction comes from a dynamical coupled-channel approach,
developed at the Excited Baryon Analysis Center (EBAC) \cite{EBAC_07}.
Six intermediate states, including $\pi$N, $\eta$N, $\pi\Delta$, $\sigma$N, $\rho$N and $\omega$N,
are used to describe the unpolarized cross sections.
Predictions for the $\Sigma$ beam asymmetry on the $\omega$ photo-production data are shown in fig. \ref{fig:sigma_proton_theory}
as dashed lines.
$t$-channel exchange contributions are not explicitly included in the calculation,
and curves tend to over-estimate our experimental asymmetry values.
\newline

%
%
\subsection{$\omega$ photo-production on the quasi free-nucleon}
\label{subsec:res_quasifree}
\begin{figure*}
 \centering
 \vspace{2cm}
 \resizebox{0.7\textwidth}{!}{\includegraphics{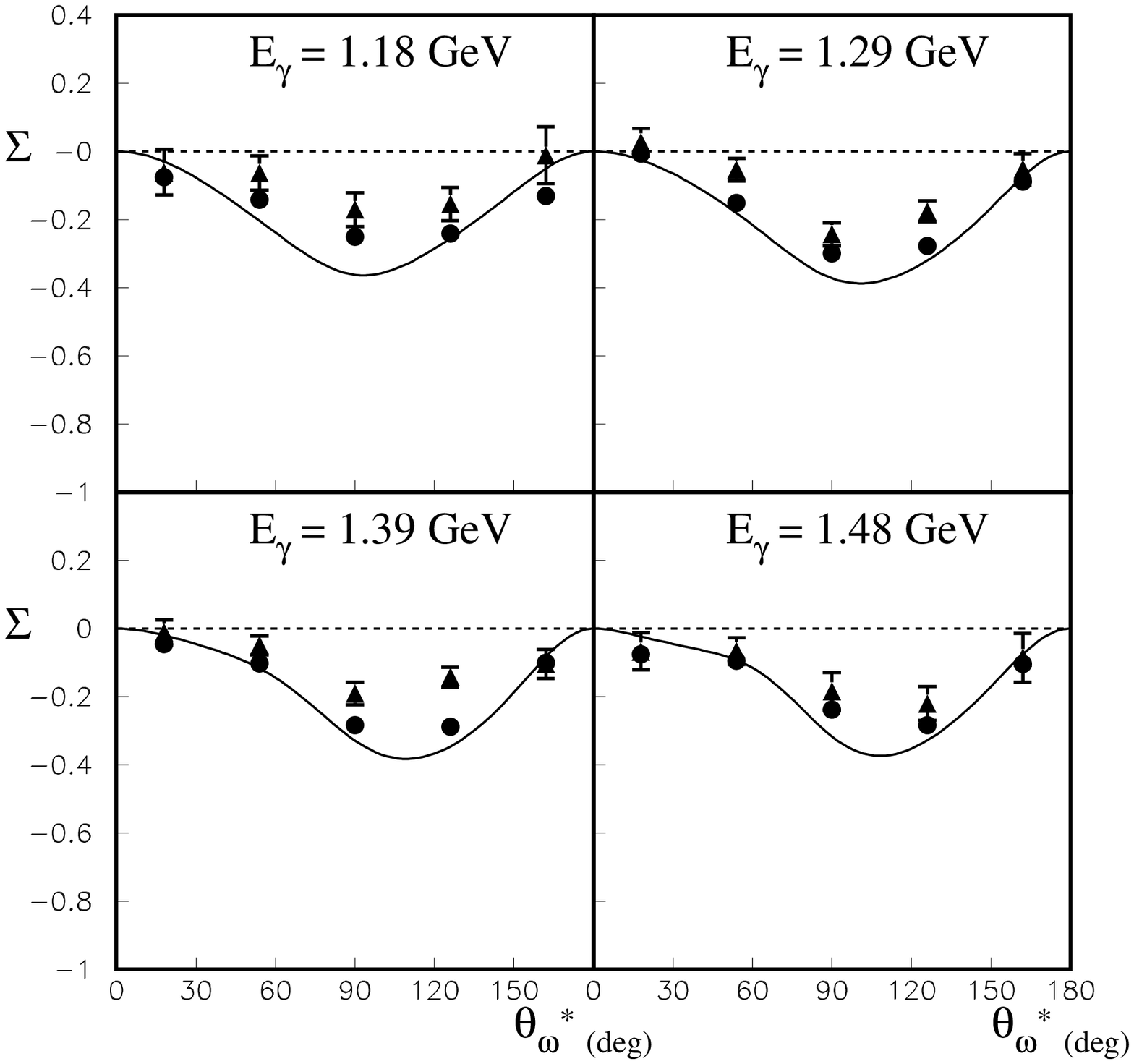}} 
 \caption{Comparison of the results obtained for the $\omega$ photo-production on the free proton (full circles)
 and on the quasi-free proton (full triangles) in a $D_2$ target.
 A generally good agreement between free proton and quasi-free proton results is observed,
 hinting at results for the quasi-free neutron as reliable to extract information
 about the $\omega$ photo-production reaction off free neutron.
 Solid curves are from \cite{Zhao_98,Zhao_01}.}
 \label{fig:sigma_qfp} 
\end{figure*}

\begin{figure*}
 \centering
 \vspace{2cm}
   \resizebox{0.7\textwidth}{!}{\includegraphics{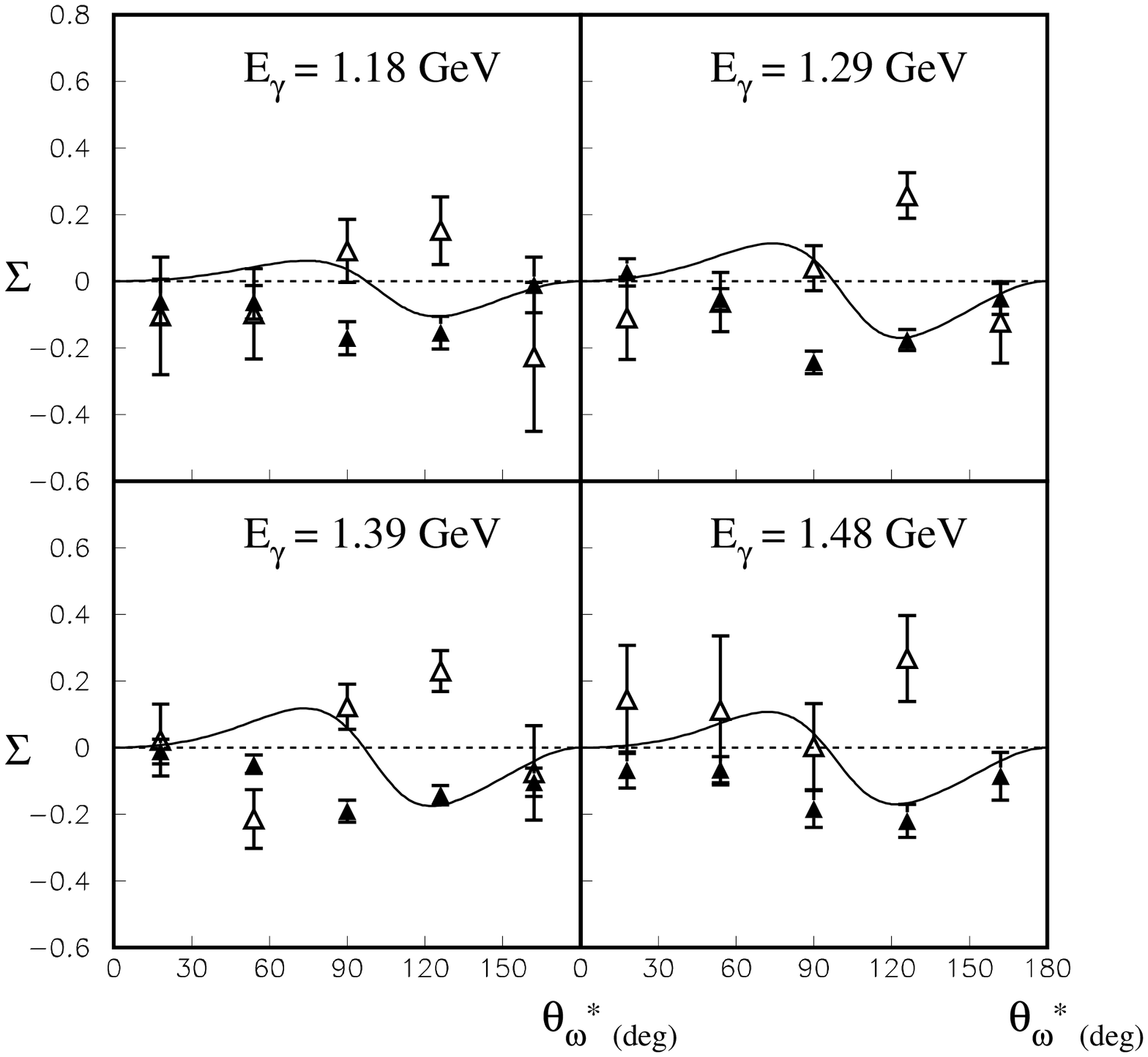}}
 \caption{Absolute first results of the $\Sigma$ beam asymmetry for the $\omega$ photo-production on the neutron (open triangles)
 from the $D_2$ target.
 They are compared with the results on the quasi-free proton (full triangles).
 Very different angular distributions are observed,
 indicating that different contributions are involved in the two reactions.
 Data are compared with the theoretical prediction from \cite{Zhao_neutron} (solid line)
 for the $\omega$ photo-production on a free neutron target.}
 \label{fig:sigma_qfn} 
\end{figure*}

$\Sigma$ beam asymmetry values are extracted for $\omega$ photo production off quasi-free proton in $D_2$ target,
from 1.1 GeV up to 1.5 GeV of incoming photon energy.
They are listed in tab. \ref{tab:results}.
Results are shown in fig. \ref{fig:sigma_qfp} as full triangles,
together with the results on the free-proton (full circles)
and with predictions from \cite{Zhao_98,Zhao_01}.
All events are obtained from the radiative decay channel analysis.
An overall good agreement is found between results on the free and the bound proton,
although a general trend of quasi-free results to be slightly lower than free results may be noticed.
Fermi motion effects may be responsible for the slight difference,
but a dedicated theoretical investigation could be useful
to better clarify the physics that lies under this small discrepancy.
Nevertheless, the generally good agreement of the two results
hints at similar conclusions about the reaction off the neutron.\\
In fig. \ref{fig:sigma_qfn} our results for the $\Sigma$ beam asymmetry
for the $\omega$ photo-production on the quasi-free neutron from a $D_2$ target are shown
(empty triangles). Numerical results are listed in tab. \ref{tab:results}, too.
These are the first results of $\Sigma$ polarization observable for the $\omega$ photo-production on the neutron.
The error bars correspond to statistical errors.
Results on the quasi-free neutron are compared with the ones on the quasi-free proton (full triangles).
The angular distribution of the beam asymmetry values differs between the neutron and proton targets,
suggesting that different reaction mechanisms could be involved in the case of $\omega$
photo-production off neutron.
For the quasi-free neutron case,
the beam asymmetry values are generally small
and in some cases compatible with zero within the error bars.
Nevertheless, the indication of a passage through zero can be observed
in all energy bins around the third data point,
being the fourth data point positive in all the energy bins.
At present, no theoretical predictions for the beam asymmetry
of the reaction $\gamma n \rightarrow \omega n$ are available in literature.
In fig. \ref{fig:sigma_qfn} our results are compared with the extension of the model
\cite{Zhao_01} to the neutron case \cite{Zhao_neutron}.
According to this model,
a change of sign in the beam asymmetry values was expected at about 90$^{\circ}$,
but our trend looks reversed.


%
%
\section{CONCLUSIONS}\label{sec:conclusion}
$\Sigma$ beam asymmetries are measured by the GRAAL collaboration
for the $\omega$ meson photo-production reaction both on the $H_2$ and the $D_2$ targets.
For the first time, results for the free proton are obtained
both for the $\omega \rightarrow \pi^0 \gamma$ and
the $\omega \rightarrow \pi^+ \pi^- \pi^0$ decay channels
from the same data set and are in strong agreement.
The results presented here provide a resolution of the existing discrepancy between previously published data.

Since the asymmetry should vanish if no resonance contributes to $\omega$ photo-production,
our experimental results strongly confirm the important role of intermediate resonant states
in the $\omega$ production mechanism.
In particular, agreement between the experimental results and the theoretical description by \cite{Zhao_01}
is observed, confirming the importance of the $P_{13}$(1720) resonance contribution to the reaction mechanism.
Different conclusions are drawn by model \cite{Ti_02},
still in agreement with our lower energy results,
but with the dominant contribution provided by the $F_{15}$(1680) resonance.

First measurements are obtained for the $\Sigma$ beam asymmetry
of the $\omega$ meson photo-production from both the quasi-free proton
and quasi-free neutron targets in the participant/spectator analysis frame.
Good agreement is found between results on the free and the quasi-free proton targets,
in analogy with the result found by our previous analysis for the $\pi^0$ and $\eta$ photo-production reactions
on the $H_2$ and $D_2$ targets \cite{DiSalvo_2009,Bartalini_2007,Fantini_2008}.
A slight trend of the quasi-free results to be lower than the free results may be noticed,
and it may be due to Fermi momentum effects.

First results for the quasi-free neutron target show that the $\Sigma$ beam asymmetry is small,
compatible with zero up to $\theta_{\omega}^* \simeq$ 90$^{\circ}$
but always positive at $\theta_{\omega}^* \simeq$ 120$^{\circ}$,
but also compatible with zero within the errors.
The angular distribution of the results for the neutron differs from the one measured for the proton target.
This suggests that different production mechanisms contribute for the two different nucleons.\\
At present, no theoretical description of beam asymmetry for
$\omega$ photo-production off neutron is available for data interpretation.

\begin{table*}
\caption{\label{tab:results} $\Sigma$ beam asymmetry values for the different reactions.
  		The $E_{\gamma}$ values (first column) correspond to the average incoming photon energy weighted by the number of corresponding photons.
                 The $\theta_{\omega}^*$ values (second column) correspond to the middle value of the $\theta_{\omega}^*$ bins.
                 The beam asymmetry values are listed together with their statistical errors.}
\begin{ruledtabular}
\begin{tabular}{|c|c|c|c|c|c|}
       \multicolumn{2}{|c|}{}                                     & \multicolumn{4}{c|}{Reaction}\\
       \hline
       $<E_{\gamma}>$ & $<\theta_{\omega}^*>$ & $\gamma p \rightarrow \omega p $   & $\gamma p \rightarrow \omega p$      & $\gamma p (n) \rightarrow \omega p (n)$ &  $\gamma n (p) \rightarrow \omega n (p)$\\
                                    &                                       & $\omega \rightarrow \pi^0 \gamma$ & $\omega \rightarrow \pi^+ \pi^0 \pi^-$ &  $\omega \rightarrow \pi^0 \gamma$         & $\omega \rightarrow \pi^0 \gamma$          \\
        \hline
          \multirow{5}{*}{1.18 GeV}&  15$^{\circ}$  & -0.076 $\pm$ 0.021                      & -0.049 $\pm$ 0.009                         & -0.06 $\pm$ 0.06                               & -0.10 $\pm$ 0.18 \\
                                                  &  54$^{\circ}$  & -0.142 $\pm$ 0.015                      & -0.155 $\pm$ 0.009                         & -0.06 $\pm$ 0.05                               & -0.01 $\pm$ 0.14 \\
                                                  &  90$^{\circ}$  & -0.250 $\pm$ 0.015                      & -0.279 $\pm$ 0.009                         & -0.17 $\pm$ 0.05                               &  0.09 $\pm$ 0.09 \\
                                                  & 126$^{\circ}$ & -0.241 $\pm$ 0.016                      & -0.274 $\pm$ 0.009                         & -0.15 $\pm$ 0.05                               &  0.15 $\pm$ 0.10 \\
                                                  & 162$^{\circ}$ & -0.132 $\pm$ 0.026                      & -0.095 $\pm$ 0.009                         & -0.001 $\pm$ 0.083                           & -0.23 $\pm$ 0.22 \\
         \hline
          \multirow{5}{*}{1.29 GeV}&  15$^{\circ}$  & -0.007 $\pm$ 0.014                      &  0.013 $\pm$ 0.008                         &  0.03 $\pm$ 0.04                               & -0.11 $\pm$ 0.12 \\
                                                  &  54$^{\circ}$  & -0.151 $\pm$ 0.011                      & -0.278 $\pm$ 0.008                         & -0.05 $\pm$ 0.03                               & -0.06 $\pm$ 0.09 \\
                                                  &  90$^{\circ}$  & -0.299 $\pm$ 0.010                      & -0.525 $\pm$ 0.008                         & -0.24 $\pm$ 0.03                               &  0.04 $\pm$ 0.07 \\
                                                  & 126$^{\circ}$ & -0.277 $\pm$ 0.010                      & -0.336 $\pm$ 0.008                         & -0.18 $\pm$ 0.03                               &  0.26 $\pm$ 0.07 \\
                                                  & 162$^{\circ}$ & -0.089 $\pm$ 0.016                      & -0.116 $\pm$ 0.008                         & -0.05 $\pm$ 0.05                               & -0.12 $\pm$ 0.12 \\
          \hline
          \multirow{5}{*}{1.39 GeV}&  15$^{\circ}$  & -0.046 $\pm$ 0.014                      & -0.014 $\pm$ 0.008                         & -0.01 $\pm$ 0.04                               &  0.02 $\pm$ 0.11 \\ 
                                                  &  54$^{\circ}$  & -0.103 $\pm$ 0.011                      & -0.068 $\pm$ 0.008                         & -0.05 $\pm$ 0.03                               & -0.21 $\pm$ 0.09 \\
                                                  &  90$^{\circ}$  & -0.284 $\pm$ 0.010                      & -0.373 $\pm$ 0.008                         & -0.19 $\pm$ 0.03                               &  0.12 $\pm$ 0.07 \\
                                                  & 126$^{\circ}$ & -0.288 $\pm$ 0.009                      & -0.371 $\pm$ 0.008                         & -0.14 $\pm$ 0.03                               &  0.23 $\pm$ 0.06 \\
                                                  & 162$^{\circ}$ & -0.100 $\pm$ 0.015                      & -0.132 $\pm$ 0.008                         & -0.10 $\pm$ 0.04                               & -0.08 $\pm$ 0.14 \\
           \hline
          \multirow{5}{*}{1.48 GeV}&  15$^{\circ}$  & -0.076 $\pm$ 0.022                      & -0.023 $\pm$ 0.008                         & -0.07 $\pm$ 0.05                               &  0.15 $\pm$ 0.16 \\
                                                  &  54$^{\circ}$  & -0.094 $\pm$ 0.016                      & -0.047 $\pm$ 0.008                         & -0.07 $\pm$ 0.04                               &  0.12 $\pm$ 0.22 \\
                                                  &  90$^{\circ}$  & -0.237 $\pm$ 0.017                      & -0.254 $\pm$ 0.008                         & -0.18 $\pm$ 0.06                               &  0.003 $\pm$ 0.129 \\
                                                  & 126$^{\circ}$ & -0.283 $\pm$ 0.014                      & -0.270 $\pm$ 0.008                         & -0.22 $\pm$ 0.05                               &  0.27 $\pm$ 0.13 \\
                                                  & 162$^{\circ}$ & -0.104 $\pm$ 0.023                      & -0.069 $\pm$ 0.008                         & -0.09 $\pm$ 0.07                               & -0.34 $\pm$ 0.52 \\
    \end{tabular}
\end{ruledtabular}
\end{table*}

We are grateful to Q. Zhao and M. Paris for the interesting and useful discussions
and for the communication of most recent results.
We thank the ESRF as a host institution for the stable operation of the storage ring.
For their support in the maintenance of the apparatus,
we thank the technical staff of the collaboration and in particular G. Nobili.

\end{document}